# Significance of interphase boundaries on activation of high-entropy alloys for room-temperature hydrogen storage

Shivam Dangwal[1,2] and Kaveh Edalati[1,2,]*

[1] WPI, International Institute for Carbon-Neutral Energy Research (WPI-I2CNER), Kyushu University, Fukuoka 819-0395, Japan

[2] Department of Automotive Science, Graduate School of Integrated Frontier Sciences, Kyushu University, Fukuoka 819-0395, Japan

The ability of high-entropy alloys (HEAs) for hydrogen storage is a rather new topic in the hydrogen community. HEAs with the C14 Laves phase have shown a high potential to reversibly store hydrogen at room temperature, but most of these alloys require a high-temperature activation treatment. This study explores the role of interphase boundaries on the easy activation of HEAs at room temperature. Two chemically similar HEAs with single and dual phases, $TiV_{1.5}ZrCr_{0.5}MnFeNi$ (C14 + 4 vol% BCC phases) and $TiV_{1.5}Zr_{1.5}CrMnFeNi$ (single C14 phase), are designed and synthesized. While the dual-phase alloy readily absorbs hydrogen at room temperature without any activation treatment, the single-phase alloy requires a high-temperature activation. It is suggested that interphase boundaries not only provide pathways for easy hydrogen transport and activation of HEAs at room temperature but also act as active sites for heterogeneous nucleation of hydride. This study introduces interphase-boundary generation as an effective strategy to address the activation drawback of HEAs.

*Corresponding author

Kaveh Edalati (E-mail: kaveh.edalati@kyudai.jp; Tel/Fax: +81 92 802 6744)

## 1. Introduction

Hydrogen is a promising fuel to meet the energy demands in the future because of its high abundance on the Earth, high energy density, and the potential of its clean production. The combustion of hydrogen produces only water vapor, which makes it an environment-friendly fuel [1]. One main issue that determines the future of hydrogen fuel is the efficiency and safety of its storage [2]. Attempts for safe and compact storage of hydrogen have led to the development of metal hydrides for storing hydrogen in the solid-state form for several decades [3]. Compared to high-pressure gas and cryogenic liquid forms, these metal hydrides have advantages such as high volumetric density, low operation pressure and simple hydrogen uptake [3]. However, the challenge with the majority of metal hydrides is that they require high temperatures for either desorption of hydrogen or activation for hydriding [4].

A recent solution concerning the high-temperature requirement for hydrogen absorption and desorption in metal hydrides is the use of high-entropy alloys (HEAs) and corresponding high-entropy hydrides [5]. It has been reported that some HEAs can perform hydrogen storage at room temperature with complete reversibility [5,6], but most of these alloys still require a high-temperature activation treatment [6-14]. To solve the activation problem of hydrogen storage materials such as Mg [15], $Mg_2Ni$ [16], TiFe [17] and TiV [18], planar defects such as grain boundaries and stacking faults have been reported to act as quick hydrogen pathways for hydrogen absorption. Moreover, it was shown that the addition of a second phase and the formation of interphase boundaries can provide preferential hydrogen pathways to ease the activation of hydrogen storage materials [19-23]. Since the limited HEAs that can absorb hydrogen without an activation process contain a second phase [5,24], one may expect that interphase boundaries can be effective for the easy activation of HEAs as well, although there has been no research in this regard.

The objective of this study is to clarify the effect of interphase boundaries on the activation of HEAs for hydrogen storage. To accomplish this, two different AB-type HEAs (A are the elements with a strong affinity for hydrogen and B are the elements with a weak affinity for hydrogen) with C14 Laves + BCC phases ($TiV_{1.5}ZrCr_{0.5}MnFeNi$) and single C14 phase ($TiV_{1.5}Zr_{1.5}CrMnFeNi$) were synthesized. Both alloys were designed by keeping the valence electron concentration (VEC) greater than 6, electronegativity higher than 7 and atomic size mismatch more than 5% [25-27]: $TiV_{1.5}ZrCr_{0.5}MnFeNi$ (VEC = 6.214, atomic size mismatch 8.64% and electronegativity difference 10.93%) and TiV$_{1.5}$Zr$_{1.5}$CrMnFeNi (VEC = 6.062, atomic size mismatch 8.82% and electronegativity difference 11.31%). It is confirmed that the presence of a second phase and corresponding interphase boundaries lead to hydrogen absorption without any extra activation process, while the single-phase HEA needs to be activated at 673 K. These results suggest a simple approach to solve the activation problem of HEAs for hydrogen storage by the introduction of interphase boundaries.

## 2. Experimental Procedures

High-purity titanium (99.9%), vanadium (99.7%), zirconium (99.2%), chromium (99.99%), manganese (99.9%), iron (99.9%) and nickel (99.99%) were used to make the HEAs TiV$_{1.5}$ZrCr$_{0.5}$MnFeNi and TiV$_{1.5}$Zr$_{1.5}$CrMnFeNi via vacuum arc melting on a water-cooled copper mold, as attempted earlier [5,24]. To achieve homogeneous compositions, each alloy was turned and remelted seven times. For microstructural characterizations, cylindrical discs with 0.8 mm thickness and 10 mm diameter were made from the ingot using a wire-cutting electric discharge machine. The grinding of the discs was done using emery papers (grit numbers: 800, 1200, 2000) followed by buffing using 9 µm and 3µm diamond suspension and final polishing using 60 nm colloidal silica. A mirror-like surface was achieved after this process.

X-ray diffraction (XRD) was used to determine the crystal structure using Cu Kα radiation with an acceleration voltage of 45 kV and a filament current of 200 mA. Phase analysis using PDXL software was employed to ascertain the phases, their lattice parameters and their fractions.

The microstructure was first studied using a scanning electron microscope (SEM) with an acceleration voltage of 15 kV. The mirror-like polished discs were used to examine the microstructure and the compositions of phases using electron backscatter diffraction (EBSD) and energy-dispersive X-ray spectroscopy (EDS). The fraction of phases present in SEM micrographs was also determined using the image segmentation tool in MATLAB. The nanostructure was studied using a transmission electron microscope (TEM) and scanning-transmission electron microscope (STEM). For TEM and STEM, the ingots were crushed in a medium of alcohol and these crushed samples were dispersed onto carbon grids. High-resolution imaging, fast Fourier transform (FFT) and EDS were utilized to study the nanostructure from TEM and STEM at 200 kV.

The hydrogen absorption-desorption pressure-composition-temperature (PCT) analysis and kinetics test were performed by a Sievert-type machine at 303 K. For hydrogen storage tests, crushing of the sample was done in the air followed by passing the crushed powders through a sieve of 75 µm size. The hydrogenation of the alloys was first conducted without any activation treatment, but the alloy that did not absorb hydrogen was further activated under vacuum at 673 K for 1 h. After the third cycle of PCT analysis at 303 K, the kinetics test was conducted for 1 h under an initial hydrogen pressure of 3.5 MPa at 303 K. The hydrogenated samples were examined by XRD in less than 5 min after their removal from the reactor of the Sievert-type machine to identify the hydride phase.

## 3. Results
### 3.1. Crystal Structures of Alloys

XRD analysis, shown in Fig. 1, confirms that TiV$_{1.5}$ZrCr$_{0.5}$MnFeNi is mainly made up of the C14 Laves phase (P63/mmc, $a = b = 0.493$ nm, $c = 0.809$ nm) with a small amount (~1 wt%) of the BCC phase (Im-3m, $a = b = c = 0.299$ nm). For TiV$_{1.5}$Zr$_{1.5}$CrMnFeNi, all the peaks observed in the XRD analysis correspond to those of the C14 Laves phase (P63/mmc, $a = b = 0.499$ nm, $c = 0.817$ nm). These XRD analyses indicate the successful design and synthesis of single- and dual-phase HEAs based on the Laves phase. Previous research has reported the ability of the C14 Laves phase to absorb hydrogen reversibly in conventional alloys [28-30] and HEAs [26,27,31].

### 3.2. Microstructure of TiV$_{1.5}$ZrCr$_{0.5}$MnFeNi

The SEM micrograph for the TiV$_{1.5}$ZrCr$_{0.5}$MnFeNi alloy is shown in Fig. 2(a). A dendritic structure was revealed by SEM, while the changes in image contrast suggest that there is a compositional gradient in the dendrites due to rapid solidification during arc melting. Close examination of the microstructure by EBSD suggests that in addition to the dendrites which have the C14 structure, there are some spherical particles with the BCC phase, as shown in the EBSD phase map of Fig. 2(b). The percentage of the C14 Laves phase is determined 96 vol% while the BCC phase has only 4 vol%. The formation of dendritic microstructure with compositional gradient is rather common and it was reported in various HEAs [32,33]. The EDS elemental mapping is shown in Fig. 2(c) and the composition of the phases and their grain sizes are summarized in Table 1. The spherical BCC phase particles are rich in vanadium and chromium, as two BCC stabilizer elements [34,35], and poor in zirconium as a Laves phase stabilizer [36].

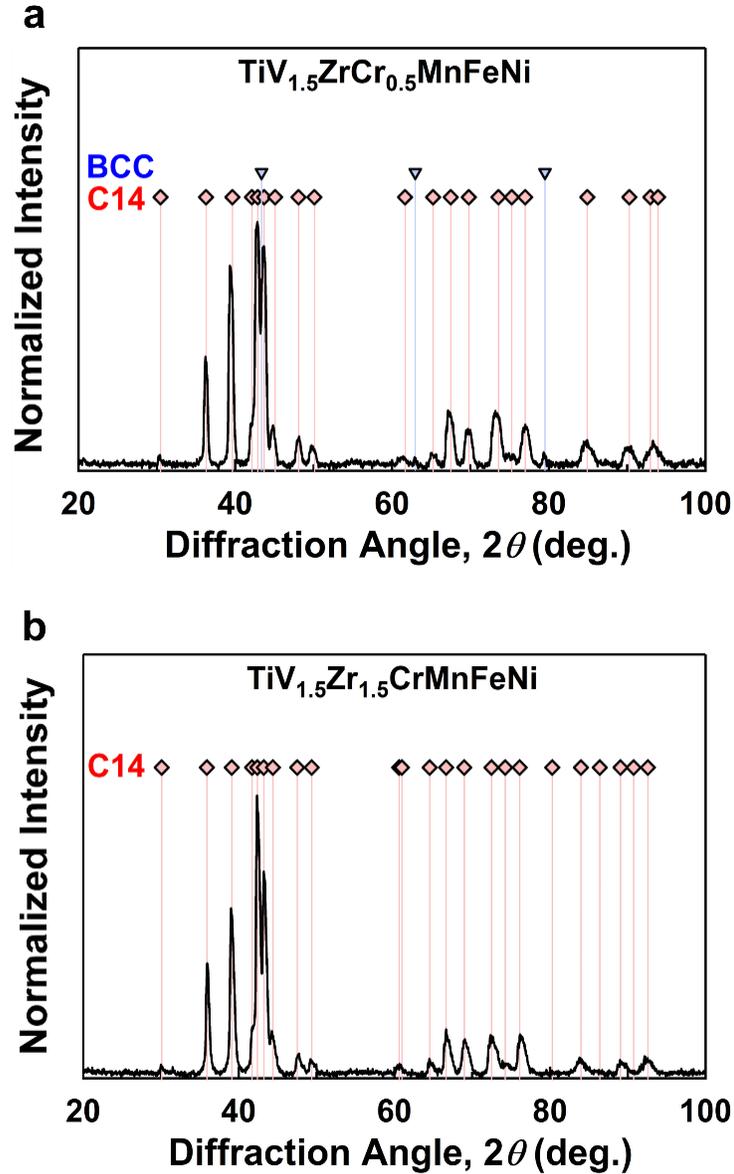

Figure 1. Producing single C14 phase and C14+BCC phases in Ti-V-Zr-Cr-Mn-Fe-Ni system. XRD profiles of (a) TiV$_{1.5}$ZrCr$_{0.5}$MnFeNi and (b) TiV$_{1.5}$Zr$_{1.5}$CrMnFeNi high-entropy alloys.

Examination of the distribution of elements in TiV$_{1.5}$ZrCr$_{0.5}$MnFeNi at the nanometer scale is shown in Fig. 3(a) using high-angle annular dark-field (HAADF) images and corresponding STEM-EDS mappings. Despite compositional gradients at the micrometer level, the distribution of elements at the nanometer level is reasonably uniform and the seven elements co-exist in the nanostructure. Examination of the nanostructure by high-resolution images confirms that the majority of the alloy contained the C14 phase, as shown in Fig. 3(b). However, the BCC phase and C14/BCC interphase boundaries can be detected easily in the microstructure, as shown in Fig. 3(c). These high-resolution analyses are in line with XRD and EBSD analyses for this alloy. All these microstructural analyses confirm the successful introduction of interphase boundaries in TiV$_{1.5}$ZrCr$_{0.5}$MnFeNi, which may act positively to enhance the activation for hydrogen storage [19-23].

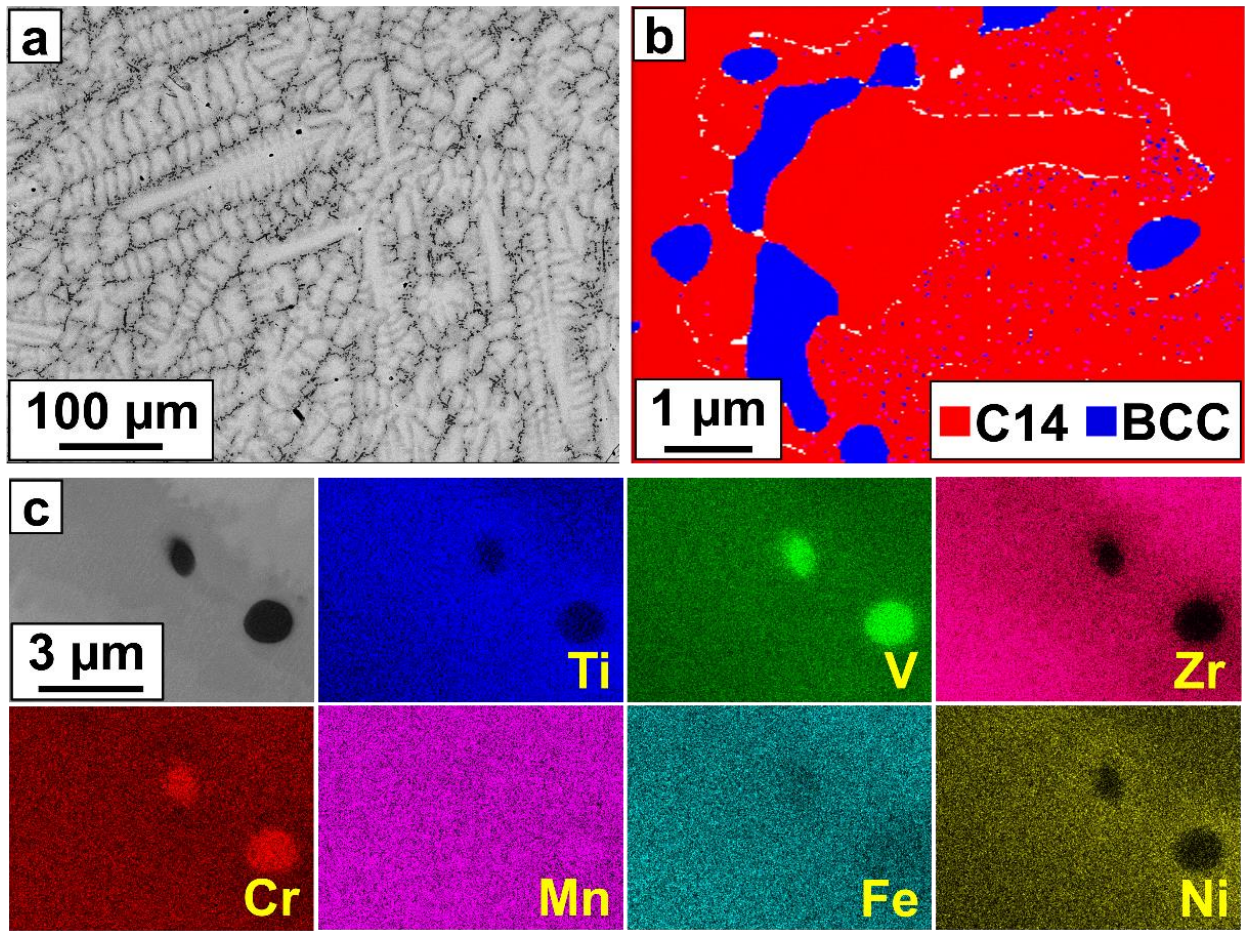

Figure 2. Formation of BCC phase between the dendrites of C14 phase in TiV$_{1.5}$ZrCr$_{0.5}$MnFeNi high-entropy alloy. (a) SEM micrograph taken by backscatter electrons, (b) EBSD phase map with beam step size of 70 nm, and (c) SEM image taken by secondary electron and corresponding EDS elemental mappings.

Table 1. Lattice parameters, phase fractions, average grain sizes, and chemical compositions of C14 and BCC phases in TiV$_{1.5}$ZrCr$_{0.5}$MnFeNi high-entropy alloy.

|  | Evaluation Method | Nominal | Overall | C14 Phase | BCC Phase |
|---|---|---|---|---|---|
| **Lattice Parameters (nm)** | XRD |  |  | $a = 0.493$, $c = 0.809$ | $a = 0.299$ |
| **Phase Fraction (wt%)** | XRD |  |  | 99 | 1 |
| **Phase Fraction (vol%)** | SEM |  |  | 96 | 4 |
| **Grain Size (μm)** | SEM |  |  | 16.6 | 1.9 |
| **Ti (at%)** | EDS | 14.3 | 14.8 | 15.2 | 6.8 |
| **V (at%)** |  | 21.4 | 21.3 | 19.8 | 53.6 |
| **Zr (at%)** |  | 14.3 | 14.9 | 15.5 | 0.3 |
| **Cr (at%)** |  | 7.1 | 6.8 | 6.6 | 10.6 |
| **Mn (at%)** |  | 14.3 | 13.9 | 13.9 | 13.9 |
| **Fe (at%)** |  | 14.3 | 13.9 | 14.2 | 10.0 |
| **Ni (at%)** |  | 14.3 | 14.4 | 14.8 | 4.8 |

### 3.3. Microstructure of TiV$_{1.5}$Zr$_{1.5}$CrMnFeNi

Fig. 4 shows the SEM micrograph, EBSD phase mapping, and EDS elemental mappings for TiV$_{1.5}$Zr$_{1.5}$CrMnFeNi. Fig. 4(a) confirms the presence of gradient composition in this alloy, as evident by image contrast changes. Examination of this alloy by EBSD, as shown in Fig. 4(b), suggests that the alloy is almost a single-phase HEA and the fraction of BCC phase is less than 0.5 vol% which is consistent with XRD analysis. The formation of a single C14 phase in this alloy is due to the high amount of zirconium which produces a high atomic size mismatch to stabilize the C14 Laves phase [36]. Fig. 4(c) shows EDS elemental mappings, which do not provide any evidence for the presence of Cr- and V-rich BCC particles. The quantitative amount of phase composition along with lattice parameters and grain size is given in Table 2. These microstructural analyses confirm that the fraction of interphase boundaries in TiV$_{1.5}$Zr$_{1.5}$CrMnFeNi is negligible compared to the dual-phase HEA TiV$_{1.5}$ZrCr$_{0.5}$MnFeNi. The design and successful synthesis of the single-phase Laves phase from seven different elements is a good finding because C14 is considered a promising phase for hydrogen storage [26-31].

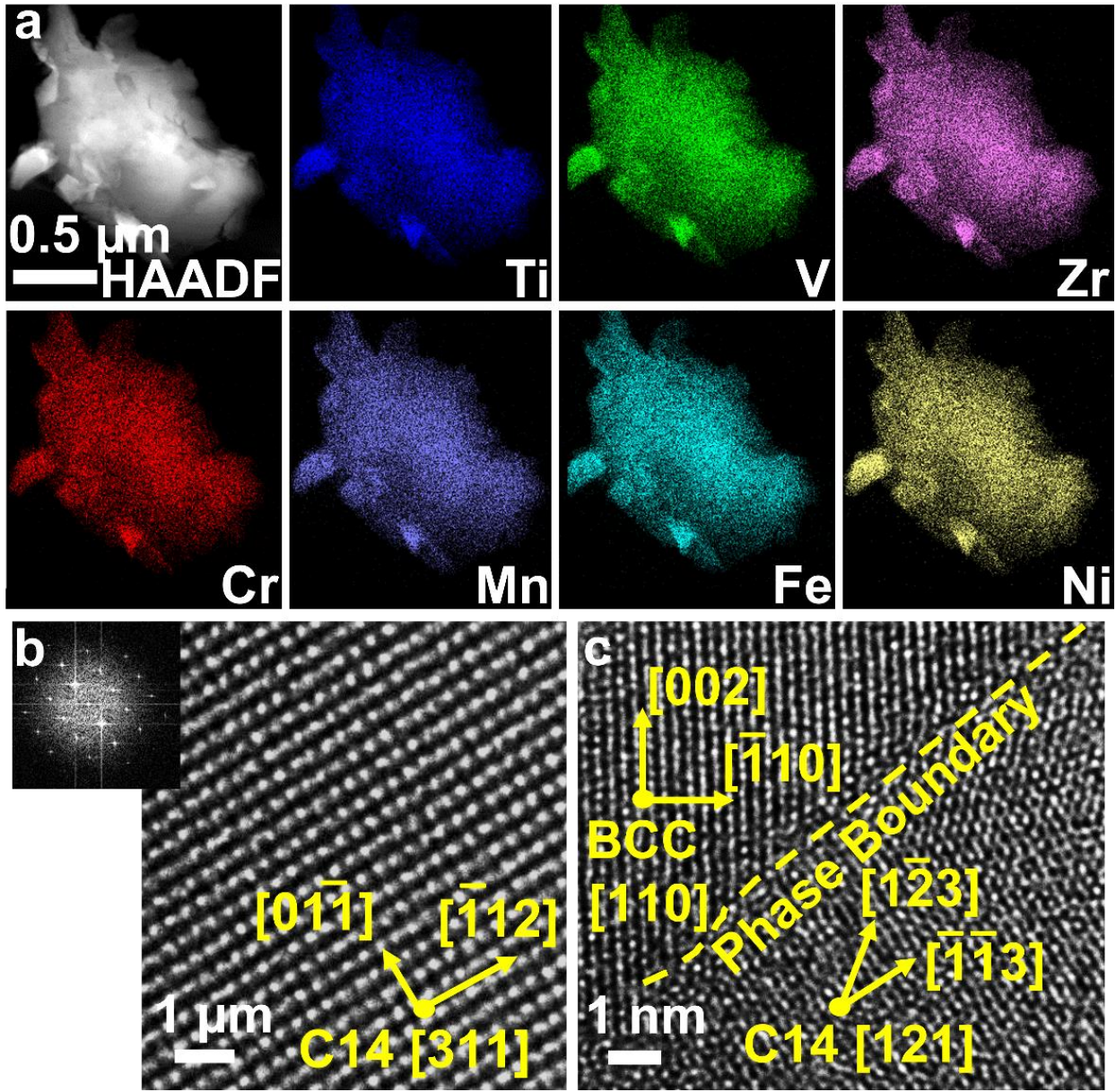

Figure 3. Formation of C14/BCC interphase boundaries in TiV$_{1.5}$ZrCr$_{0.5}$MnFeNi high-entropy alloy. (a) HAADF micrograph and corresponding STEM-EDS elemental mappings, (b) high-resolution TEM image of grain interior and corresponding fast Fourier transform and (c) high-resolution TEM image of interphase boundary.

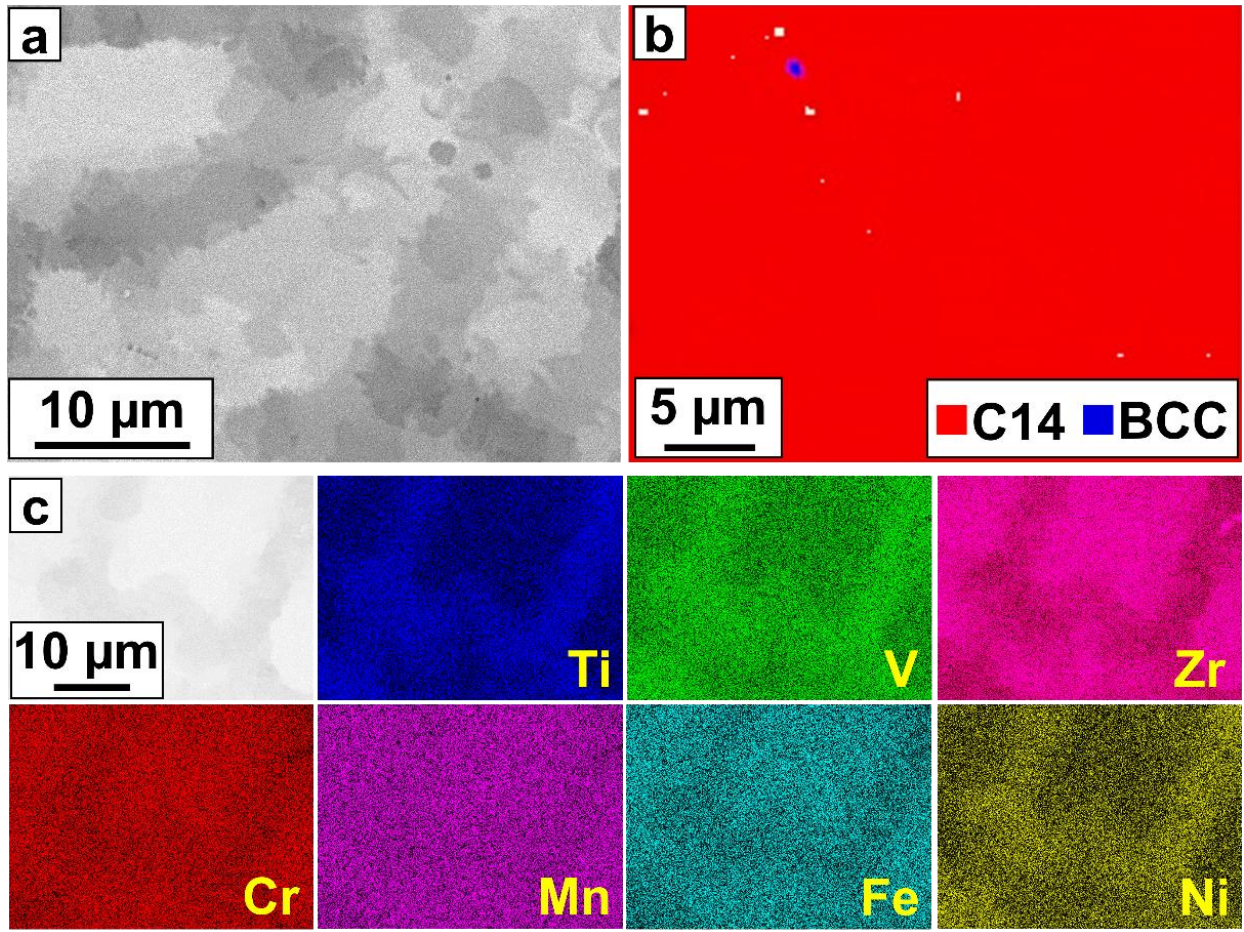

Figure 4. Formation of single C14 phase with composition gradient in TiV$_{1.5}$Zr$_{1.5}$CrMnFeNi high-entropy alloy. (a) SEM micrograph taken by backscatter electrons, (b) EBSD phase map with beam step size of 70 nm, and (c) SEM image taken by secondary electron and corresponding EDS elemental mappings.

Table 2. Lattice parameters, phase fractions, average grain sizes, and chemical compositions of C14 and BCC phases in high-entropy alloy TiV$_{1.5}$Zr$_{1.5}$CrMnFeNi.

|  | Evaluation Method | Nominal | Overall | C14 Phase | BCC Phase |
|---|---|---|---|---|---|
| **Lattice Parameters (nm)** | XRD |  |  | $a = 0.499, c = 0.817$ | _ |
| **Phase Fraction (wt%)** | XRD |  |  | 100 | 0 |
| **Phase Fraction (vol%)** | SEM |  |  | 99 | <0.5 |
| **Grain Size** (µm) | SEM |  |  | 11.1 | 1.8 |
| **Ti (at%)** | EDS | 12.5 | 12.8 | 12.8 | 9.1 |
| **V (at%)** |  | 18.75 | 19.0 | 18.9 | 49.2 |
| **Zr (at%)** |  | 18.75 | 19.3 | 19.5 | 2.6 |
| **Cr (at%)** |  | 12.5 | 12.6 | 12.6 | 16.9 |
| **Mn (at%)** |  | 12.5 | 11.3 | 11.2 | 9.7 |
| **Fe (at%)** |  | 12.5 | 12.3 | 12.3 | 7.1 |
| **Ni (at%)** |  | 12.5 | 12.7 | 12.7 | 5.5 |

Fig. 5(a) shows the HAADF image and corresponding STEM-EDS mappings for TiV$_{1.5}$Zr$_{1.5}$CrMnFeNi. The distribution of seven elements is uniform at the nanometer scale despite their heterogenous distribution at larger scales. Examination of the nanostructure by several high-resolution images could detect only the C14 phase, as shown in Fig. 5(b) and 5(c). All these structural and microstructural analyses confirm that the fraction of interphase boundaries in TiV$_{1.5}$Zr$_{1.5}$CrMnFeNi alloy is too low to influence its hydrogenation properties [19-23]. Moreover, due to the large grain size of the C14 phase (~1.1 µm), the fraction of grain boundaries is not also high enough to influence the activation for hydrogen storage [15-18].

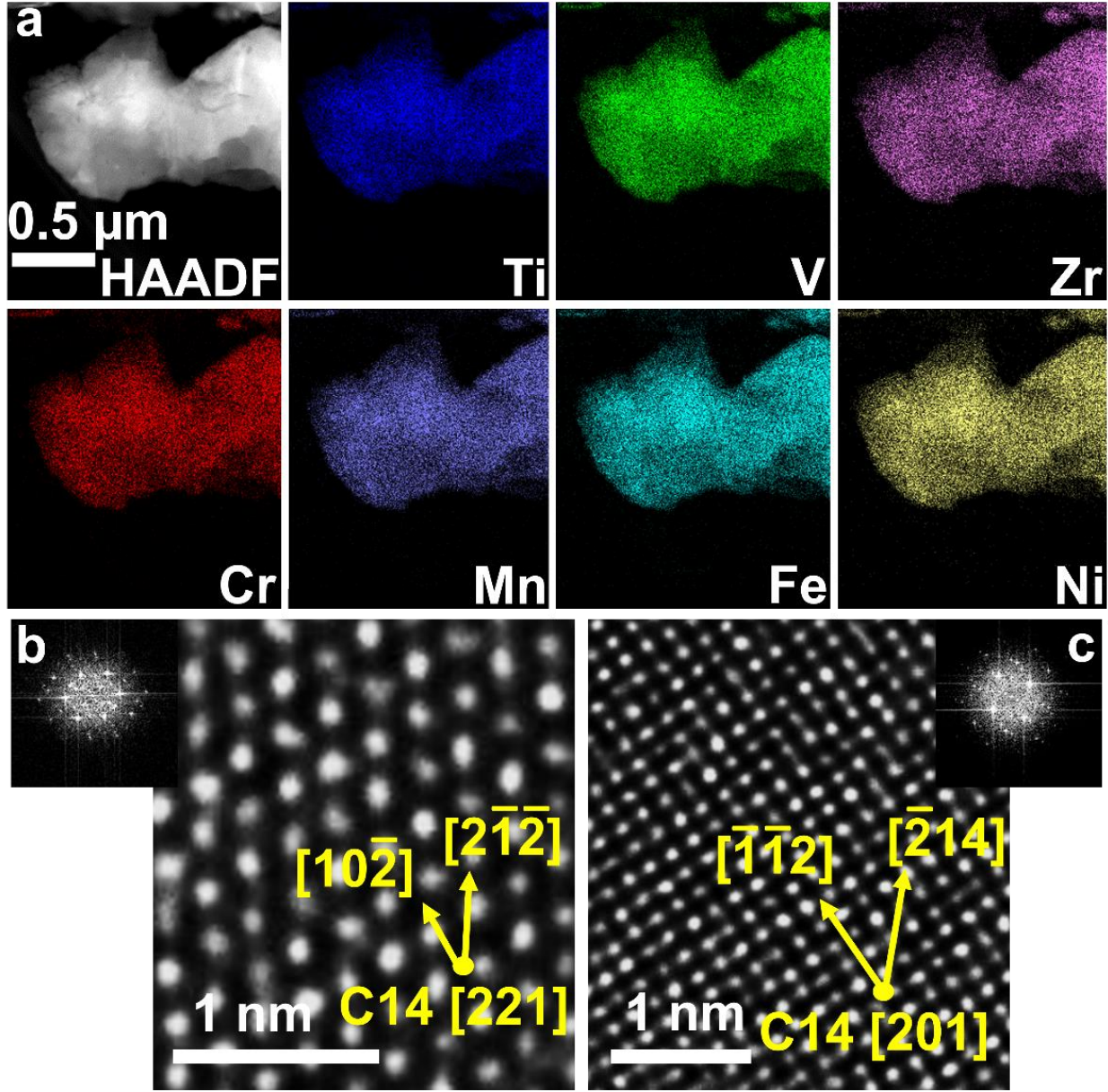

Figure 5. Formation of C14 phase in TiV$_{1.5}$Zr$_{1.5}$CrMnFeNi high-entropy alloy. (a) HAADF micrograph and corresponding STEM-EDS elemental mappings, (b, c) high-resolution TEM image of grain interior and corresponding fast Fourier transforms.

### 3.4. Hydrogen storage of alloys

The PCT isotherms for the dual-phase TiV$_{1.5}$ZrCr$_{0.5}$MnFeNi alloy are shown in Fig. 6(a). The alloy absorbs hydrogen without any activation process at room temperature. The amount of hydrogen absorbed by the alloy is 1.2-1.3 wt% (hydrogen-to-metal atomic ratio of $H/M$ = 0.7-0.8) and the alloy shows good reversibility. The kinetics test, as shown in Fig. 6(b), confirms that the hydrogen absorption speed is fast, and hydrogenation occurs in a few minutes. Examination of the hydrogenated sample after the kinetics test by XRD is shown in Fig. 6(c). It is evident that the BCC phase remains unchanged after hydrogenation, indicating that BCC does not contribute to the process as a storing media. However, the C14 phase transforms into an expanded C14 phase (P63/mmc, $a = b = 0.512$ nm, $c = 0.850$ nm), indicating that the hydrogen is stored in the Laves phase, leading to 13.3 vol% expansion. The formation of the Laves phase hydride with a $H/M$ ratio close to one is usually expected in HEAs after full hydrogenation [5,24,26,27].

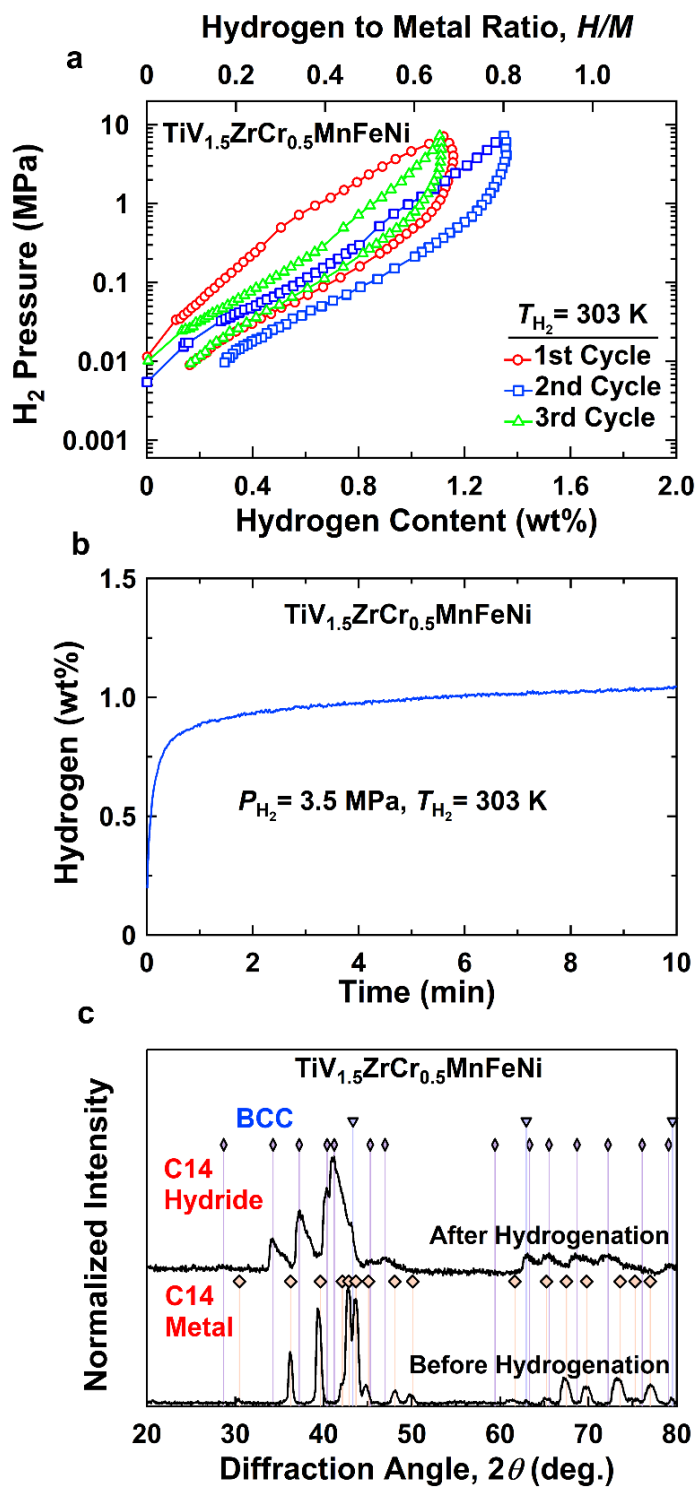

Figure 6. Fast and reversible hydrogen storage at room temperature in dual-phase TiV$_{1.5}$ZrCr$_{0.5}$MnFeNi high-entropy alloy without activation treatment. (a) PCT absorption/desorption isotherms at 303 K, (b) hydrogenation kinetic curve at 303 K, and (c) XRD profile before and after hydrogenation.

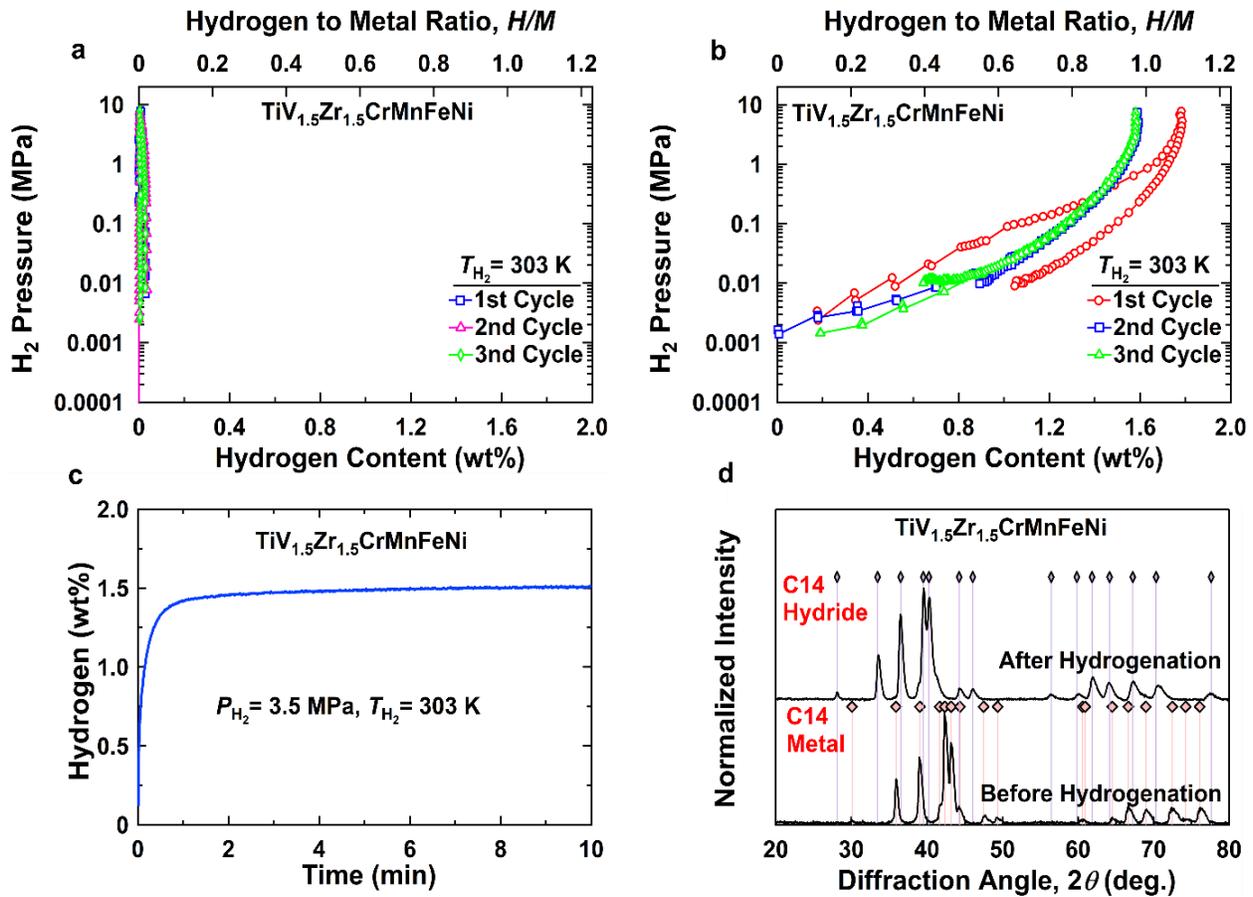

Figure 7. Fast and reversible hydrogen storage at room temperature in dual-phase TiV$_{1.5}$Zr$_{1.5}$CrMnFeNi high-entropy alloy after activation treatment at 673 K for 1 h. (a) PCT absorption/desorption isotherms at 303 K without activation treatment, (b) PCT absorption/desorption isotherms at 303 K after activation treatment at 673 K for 1 h, (c) hydrogenation kinetic curve at 303 K, and (d) XRD profile before and after hydrogenation.

Hydrogen storage results are shown in Fig. 7 for the single-phase TiV$_{1.5}$Zr$_{1.5}$CrMnFeNi alloy. The alloy does not absorb any hydrogen at room temperature without activation, as shown in Fig. 7(a). After activation at 673 K for 1 h, it started to absorb hydrogen with a capacity of 1.6-1.8 wt% ($H/M = 1.0$-1.1) as shown in Fig. 7(b). The kinetics test for this alloy is shown in Fig. 7(c), indicating that the activated material has a fast absorption kinetics. The fast kinetics in the HEA should be due to the high atomic size mismatch in the alloys which introduces lattice distortion in the structure and accordingly increases the nucleation rate of hydride [37]. Moreover, the C14 Laves phase has generally a large free volume for the transport of hydrogen and rapid kinetics [31,,38]. XRD analysis after the hydrogenation in Fig. 7(d) confirms that hydrogenation leads to a lattice expansion of 21 vol% in the Laves phase (P63/mmc, $a = b = 0.532$ nm, $c = 0.873$ nm). Since the lattice parameters of this alloy are larger than the dual-phase alloy and its VEC is lower, its higher storage capacity is reasonable [39]. However, the main difference is that the single-phase alloy does not absorb hydrogen without an activation process, a fact that can be attributed to the presence of interphase boundaries as effective planar defects for hydrogen diffusion [40,41].

## 4. Discussion

Difficult activation is a general drawback of many room-temperature hydrogen storage materials such as TiFe intermetallics and Ti-V-based alloys [7,23,41]. Difficult activation implies although these materials should thermodynamically absorb hydrogen at room temperature, they do not absorb hydrogen after exposure to hydrogen because of kinetic drawbacks [3,4]. It is believed that the difficult activation is due to the formation of an oxide layer on the surface which makes the dissociation of hydrogen and its diffusion from the bulk to the surface difficult [17]. Different methods are used to activate these materials. (i) They are usually activated by treatment at high temperatures under a vacuum or a high pressure of hydrogen [3,4,7,23]. The mechanism of activation by heating is still under argument. While some people believe a partial reduction or dissolution of oxide occurs during the process to provide fresh pathways for hydrogen, others suggested the formation of active catalysts on the surface is responsible for such activation [17]. (ii) Mechanical treatment is another activation treatment that is conducted by ball milling [15,40], cold rolling [20], severe plastic deformation [18,42] or a combination of ball milling and severe plastic deformation [43,44]. It was shown that planar defects, particularly grain boundaries and stacking faults, formed by mechanical treatment act as pathways for hydrogen transport from the surface to bulk [16,17,22]. The problem with this method is the need for an additional processing step for practical applications. (iii) The third strategy is a chemical modification by adding other elements to the alloy, which can modify either the thermodynamics by controlling stable or metastable phases or improve the kinetics by providing catalytic effects [21,23,45]. This third method is the most practical strategy and the addition of manganese to TiFe is a successful example that is commercially used [23].

HEAs and particularly those with the C14 Laves phase structure are perhaps the newest type of hydrogen storage materials that can even function at room temperature, but their activation problem should be solved [5-14]. This study attempts to find a solution to solve the activation problem of HEAs without conducting an extra high-temperature or mechanical activation treatment. Motivated by earlier studies on the efficiency of grain boundaries on activation [15-18] as well as the effect of interphase boundaries on the easy transportation of hydrogen [19-23], interphase boundaries are suggested to be added to HEAs to activate them. It is found that while single-phase HEA cannot absorb hydrogen without a high-temperature activation treatment, the dual-phase alloy can absorb hydrogen without any activation treatment. Such easy activation cannot be attributed to the difference in crystal structure because both alloys have C14 Laves phase structure. Moreover, the single-phase alloy, which is not active, has higher lattice parameters and lower VEC which are expected to ease its activation [40]. The easy activation cannot be attributed to hydrogen transport through the V-rich BCC phase because these alloys are known to suffer from poor activation despite their high affinity with hydrogen [46-48]. Moreover, the examination of hydrogenated alloys by XRD in Fig. 6 shows that hydrogen does not store in the BCC phase. The easy activation of dual-phase alloy should be due to the effect of interphase boundaries on easy hydrogen transport, a fact that was also reported in some conventional alloys and composites [15-23].

It should be noted that interphase boundaries not only act as a carrier medium for the transport of hydrogen from the surface to bulk but also amplify the heterogeneous nucleation of hydride [49]. The absorption of hydrogen takes place by the nucleation of hydride at grain boundaries, and this reduces the grain boundary energy [49]. The interphase boundaries have usually energy levels between the grain boundary energies of the individual phases [50] and the formation of hydrides with sizes larger than a critical size at these boundaries is energetically favorable to reduce the energy of the system. It is also known from classic phase transformation principles that such heterogenous nucleation needs less activation energy compared to homogeneous nucleation in the absence of a second phase [51]. Taken altogether, the addition of a small amount of a second phase appears to be a practical solution to solve the activation problem of HEAs mainly due to the effectiveness of interphase boundaries on hydrogen transport, as schematically shown in Fig. 8. This study not only introduces a new strategy for the activation of HEAs but also suggests that interphase boundaries are likely responsible for the high activity of some high-entropy hydrogen storage materials reported earlier [5,24,38]. By considering the growing significance of hydrogen storage [1] and by considering the fast development of HEAs for storing hydrogen [5-14,52-58], the current study can contribute to the design of new high-entropy hydrogen storage materials. Despite these positive findings, future studies using in-situ synchrotron and neutron diffraction methods are required to clarify the exact effect of interphase boundaries on the process of hydrogen storage from diffusion to nucleation and growth.

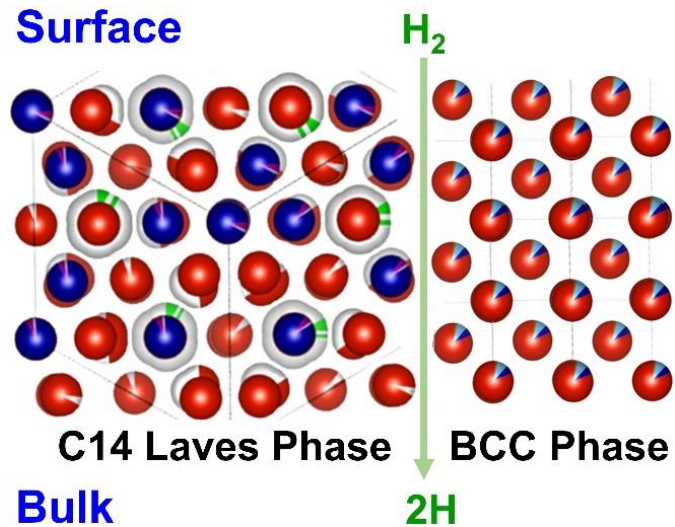

Figure 8. Schematic illustration of interface boundary and their significance on hydrogen transport from surface to bulk for easy activation for hydrogen storage.

## 5. Conclusion

This study reports the significant effect of interphase boundaries on the activation of high-entropy alloys (HEAs) for hydrogen storage at room temperature. It is shown that while a TiV$_{1.5}$Zr$_{1.5}$CrMnFeNi alloy with a single C14 phase does absorb hydrogen without a high-temperature activation, the presence of small amounts of BCC in dual-phase TiV$_{1.5}$ZrCr$_{0.5}$MnFeNi leads to its hydrogenation at room temperature without any activation problem. It is suggested that interphase boundaries mainly act as hydrogen pathways from the surface to bulk and partly as active sites for the heterogeneous nucleation of hydride. These findings introduce a rational strategy to develop active HEAs for room-temperature hydrogen storage.

**Acknowledgment**

The author S.D. thanks the MEXT, Japan for a scholarship. This work is supported in part by grants-in-aid for scientific research from the MEXT, Japan (JP19H05176 & JP21H00150).

[8]  Kunce I, Polanski M, Bystrzycki J. Microstructure and hydrogen storage properties of a TiZrNbMoV high entropy alloy synthesized using laser engineered net shaping (lens). Int J Hydrog Energy 2014;39:9904-10. https://doi.org/10.1016/j.ijhydene.2014.02.067.

[9]  Ek G, Nygård MM, Pavan AF, Montero J, Henry PF, Sørby MH, Witman M, Stavila V, Zlotea C, Hauback BC, Sahlberg M. Elucidating the effects of the composition on hydrogen sorption in TiVZrNbHf-based high-entropy alloys. Inorg Chem 2021;60:1124-32. https://doi.org/10.1021/acs.inorgchem.0c03270.

[10] Sleiman S, Huot J. Effect of particle size, pressure and temperature on the activation process of hydrogen absorption in TiVZrHfNb high entropy alloy. J Alloys Compd 2021;861:158615. https://doi.org/10.1016/j.jallcom.2021.158615.

[11] Montero J, Ek G, Laversenne L, Nassif V, Zepon G, Sahlberg M, Zlotea C. Hydrogen storage properties of the refractory Ti-V-Zr-Nb-Ta multi-principal element alloy. J Alloys Compd 2020;835:155376. https://doi.org/10.1016/j.jallcom.2020.155376.

[12] Shen H, Zhang J, Hu J, Zhang J, Mao Y, Xiao H, Zhou X, Zu X. A novel TiZrHfMoNb high-entropy alloy for solar thermal energy storage. Nanomater 2019;9:248. https://doi.org/10.3390/nano9020248.

[13] Zlotea C, Sow MA, Ek G, Couzinié JP, Perrière L, Guillot I, Bourgon J, Møller KT, Jensen TR, Akiba E, Sahlberg M. Hydrogen sorption in TiZrNbHfTa high entropy alloy. J Alloys Compd 2019;775:667-74. https://doi.org/10.1016/j.jallcom.2018.10.108.

[14] Shen H, Hu J, Li P, Huang G, Zhang J, Zhang J, Mao Y, Xiao H, Zhou X, Zu X, Long X, Peng S. Compositional dependence of hydrogenation performance of Ti-Zr-Hf-Mo-Nb high-entropy alloys for hydrogen/tritium storage. J Mater Sci Technol 2020;55:116-25. https://doi.org/10.1016/j.jmst.2019.08.060.

[15] Zaluska A, Zaluski L, Olsen JOS. Nanocrystalline magnesium for hydrogen storage. J Alloys Compd 1999;288:217-25. https://doi.org/10.1016/S0925-8388(99)00073-0.

[16] Hongo T, Edalati K, Arita M, Matsuda J, Akiba E, Horita Z. Significance of grain boundaries and stacking faults on hydrogen storage properties of $Mg_2Ni$ intermetallics processed by high-pressure torsion. Acta Mater 2015;92:46-54. https://doi.org/10.1016/j.actamat.2015.03.036.

[17] Edalati K, Akiba E, Horita Z. High-pressure torsion for new hydrogen storage materials. Sci Technol Adv Mater 2018;19:185-93. https://doi.org/10.1080/14686996.2018.1435131.

[18] Edalati K, Shao H, Emami H, Iwaoka H, Akiba E, Horita Z. Activation of titanium-vanadium alloy for hydrogen storage by introduction of nanograins and edge dislocations using high-pressure torsion. Int J Hydrog Energy 2016;41:8917-24. https://doi.org/10.1016/j.ijhydene.2016.03.146.

[19] Pinkerton FE, Meyer MS, Meisner GP, Balogh MP, Vajo JJ. Phase boundaries and reversibility of $LiBH_4$/$MgH_2$ hydrogen storage material. J Phys Chem C 2007;111:12881-5. https://doi.org/10.1021/jp0742867.
18